\documentclass[11pt,superscriptaddress]{revtex4-1}

\usepackage{graphicx}
\usepackage{dcolumn}
\usepackage{bm}
\usepackage{psfrag}
\usepackage{epsfig}
\usepackage{amsmath}
\usepackage{amssymb}
\usepackage{color}
\usepackage{bbm}
\usepackage[FIGTOPCAP,raggedright,nooneline]{subfigure}

\usepackage[colorlinks=true,citecolor=blue,linkcolor=blue,urlcolor=blue]{hyperref}

\newcommand{\ignore}[1]{}

\begin{document}

\title{$\mathcal{PT}$-symmetry breaking in the steady state of microscopic gain-loss systems}

\author{Kosmas~V.~Kepesidis$^1$, Thomas~J.~Milburn$^1$, Julian Huber$^1$, Konstantinos~G.~Makris$^2$, Stefan~Rotter$^3$ and Peter~Rabl$^1$}

\address{$^1$Vienna Center for Quantum Science and Technology,
Atominstitut, TU Wien, 1020 Vienna, Austria}
\address{$^2$Crete Center for Quantum Complexity and Nanotechnology, Department of Physics, University of Crete, 71003, Heraklion, Greece}
\address{$^3$Institute for Theoretical Physics, TU Wien, 1040 Vienna, Austria}


\begin{abstract}
The phenomenon of $\mathcal{PT}$ (parity- and time-reversal) symmetry breaking is conventionally associated with a change in the complex mode spectrum of a non-Hermitian system that marks a transition from a purely oscillatory to an exponentially amplified dynamical regime. In this work we describe a new type of $\mathcal{PT}$-symmetry breaking, which occurs in the steady-state energy distribution of open systems with balanced gain and loss. In particular, we show that the combination of nonlinear saturation effects and the presence of thermal or quantum noise in actual experiments results in unexpected behavior that differs significantly from the usual dynamical picture. We observe additional phases with preserved or `weakly' broken $\mathcal{PT}$ symmetry, and an unconventional transition from a high-noise thermal state to a low-amplitude lasing state with broken symmetry and strongly reduced fluctuations.  We illustrate these effects here for the specific example of coupled mechanical resonators with optically-induced loss and gain, but  the described mechanisms will be essential for a general understanding of the steady-state properties of actual $\mathcal{PT}$-symmetric systems operated at low amplitudes or close to the quantum regime.
\end{abstract}

%

%

%
\maketitle
%
%

\section{Introduction}

In 1998 Bender and Boettcher~\cite{BenderPRL1998} described a class of non-Hermitian `Hamiltonians' that exhibit a purely real energy spectrum, a surprising fact which they attributed to the underlying $\mathcal{PT}$ (parity- and time-reversal) symmetry. Their observation triggered considerable interest in discrete and continuous systems with $\mathcal{PT}$ symmetry along with alternative non-Hermitian formulations of quantum theory~\cite{Bender2007}. Such fundamental considerations remain speculative, but there exist many classical systems in which $\mathcal{PT}$-symmetric dynamics can be obtained with appropriately engineered loss and gain.  This was first pointed out in the context of photonic waveguides~\cite{Makris2008, Musslimani2008, Guo2009, Rueter2010}, lattices, and resonators~\cite{Regensburger2012, Peng2014, Feng2014, Hodaei2014}.  Other examples include cold atoms~\cite{Hang2013, Haag2014, Lee2014} and optomechanical devices~\cite{Jing2014, Xu2014, Lu2015}.
Of particular interest in such systems is the $\it{breaking}$ of $\mathcal{PT}$ symmetry, i.e., when by tuning a parameter the energy spectrum becomes complex and the eigenvectors no longer exhibit the underlying $\mathcal{PT}$ symmetry. This phenomenon was first experimentally observed in optical waveguides~\cite{Guo2009, Rueter2010}, and is currently the subject of intense experimental and theoretical research.

\begin{figure}
  \centering
    \includegraphics[width=0.7\textwidth]{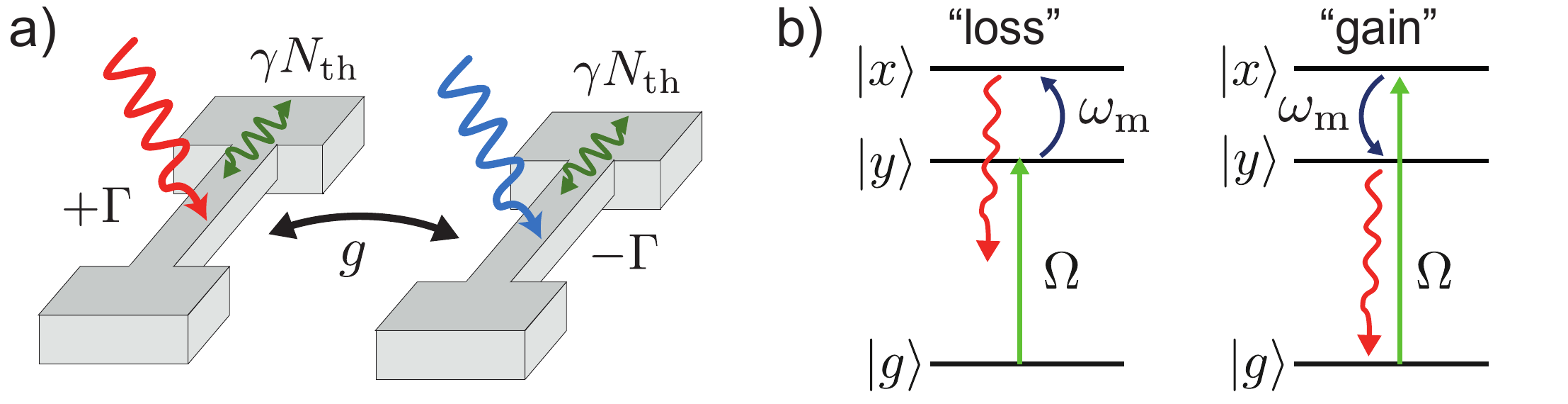}
      \caption{a)~Setup of two coupled mechanical resonators with optically-induced gain and loss. b)~Scheme for engineering mechanical gain or loss via an optically-driven three-level defect. Depending on the detuning of the pump (green arrow), phonon-induced transitions between the two near-degenerate excited states $|x\rangle$ and $|y\rangle$ lead to a net absorption or emission of phonons of frequency $\omega_{\rm m}$ (see reference~\cite{Kepesidis} for more details). }
      \label{introFig}
\end{figure}

In this work we go beyond this dynamical picture and address an interesting and still open question: what are the \emph{steady} states of actual physical systems with $\mathcal{PT}$ symmetry? 
This question becomes especially important for atomic or microscopic solid-state realizations of gain-loss systems. Here nonlinear saturation effects as well as the presence of thermal and quantum noise have a crucial influence on the system's dynamics and the long-time behavior can no longer be inferred from an eigenvalue analysis only.  By focusing on the experimentally relevant example of coupled mechanical resonators with optically-induced gain and loss (see Fig.~\ref{introFig}) we show that $\mathcal{PT}$-symmetry breaking in the steady state exhibits various unexpected features and in general occurs via additional intermediate phases with retained or `weakly' broken $\mathcal{PT}$ symmetry. Most importantly, we identify an unconventional transition from a high-noise balanced energy distribution to a parity-broken lasing state with strongly reduced fluctuations. This transition generalizes the phenomenon of $\mathcal{PT}$-symmetry breaking---hitherto defined only for eigenstates---to steady-state distributions of noisy systems. 
The mechanisms described here occur in systems of two coupled modes as well as in multi-resonator arrays, and will thus be of relevance for a large range of $\mathcal{PT}$ symmetric systems operated at low amplitudes and close to the quantum regime.

\section{Model}
To motivate the following analysis by a concrete physical system, we consider a setup of two micromechanical resonators as shown in figure~\ref{introFig}~a). The main findings discussed below, however, are more general and can be studied in other equivalent  realizations, e.g., with coupled optical or microwave resonators.  The mechanical resonators have a bare vibrational frequency $\omega_{\rm m}$ and they are mutually coupled, e.g., mechanically via the support, with strength $g$. In addition, optical or electrical cooling~\cite{WilsonRaePRL2004, MartinPRB2003, MarquardtPRL2007, WislonRaePRL2007, Kepesidis} and pumping~\cite{Kepesidis, Bennett, Rodrigues, Hauss, Vahala, Grudinin, Kabuss, Mahboob} schemes are used
to induce mechanical loss for one resonator and an \emph{equal} amount of mechanical gain for the other. In a frame rotating with $\omega_{\rm m}$, the semiclassical dynamics of the system is then described by the It\^{o} stochastic differential equation~\cite{WallsMilburn,Gardiner}
\begin{equation}\begin{split} \label{stochasticEq}
\left( \begin{array}{c}
  \dot\alpha \\
\dot \beta  \end{array} \right) 
= & 
\left( \begin{array}{cc}
 \Gamma_+(\alpha)  & -i g  \\
-ig &  \Gamma_-(\beta)  \end{array} \right)
\left( \begin{array}{c}
  \alpha \\
\beta  \end{array} \right)  
+\left( \begin{array}{c}
 F_+(t) \\
F_-(t) \end{array} \right), 
\end{split}\end{equation}
where $\alpha$ and $\beta$ are the dimensionless amplitudes of the pumped and cooled mode respectively (more details on the derivation of equation~\eqref{stochasticEq} are given in~\ref{sec:LaserTheory}).

The optically-induced gain and loss rates considered here are of the form 
\begin{equation}\label{nonlinearity}
 \Gamma_\pm (\alpha)=  \pm\frac{\Gamma}{(1 + |\alpha|^2/n_0)^\nu}-  \gamma,
\end{equation}
where $\Gamma$ is the maximal rate and $\sqrt{n_0}$ is the saturation amplitude. The value of $\nu$ characterizes the underlying heating or cooling mechanism and will be treated here as an adjustable parameter. For the three-level scheme depicted in figure~\ref{introFig}~b) this parameter takes the value $\nu=2$~\cite{Kepesidis}. Instead, for conventional laser amplification with inverted two-level systems we would obtain $\nu=1$~\cite{WallsMilburn}. Finally, $\gamma$ is the bare mechanical damping rate. Since we are interested in the $\mathcal{PT}$-symmetric limit (defined below), we assume $\gamma/\Gamma\rightarrow 0$.  However, in all our calculations we retain a finite  $\gamma > 0$, which describes the actual physical situation and results in a well-defined steady state for all parameter regimes.

In equation~\eqref{stochasticEq} the (complex) stochastic forces $F_\pm(t)$ represent two independent white-noise processes with  
$\langle F_\pm^*(t)F_\pm (t')\rangle=D_\pm \delta (t-t')$. For resonators coupled to a reservoir of temperature $T$ the diffusion rates are $D_+(\alpha) = D_{\rm q}(\alpha) + 2\gamma N_{\rm th}$ and $D_-=  2\gamma N_{\rm th}$, where $N_{\rm th}=(e^{\hbar \omega_{\rm m}/k_BT}-1)^{-1}$.  The contribution $D_{\rm q}(\alpha\rightarrow 0)=2 \Gamma$ for the gain mode represents the intrinsic quantum noise associated with any amplification process and suggests that noise is a fundamental property of $\mathcal{PT}$-symmetric systems~\cite{Schomerus2010, AgarwalPRA2012, He2014}.


\section{$\mathcal{PT}$-symmetry breaking}

By ignoring the effect of noise and assuming $|\alpha|^2, |\beta|^2\ll n_0$, equation~\eqref{stochasticEq} can be written as  $\partial_t \psi = -i H \psi$, with a state vector  $\psi= (\alpha, \beta)^{T}$ and a non-Hermitian `Hamiltonian'
\begin{equation}
H = \left( \begin{array}{ccc}
  i\Gamma  & g  \\
g &  - i\Gamma \end{array} \right).
\end{equation}
This Hamiltonian is invariant under a combined parity ($\mathcal{P}$) and time-reversal ($\mathcal{T}$) symmetry for \emph{all} values of $ \Gamma/g$, where $\mathcal{P}:(\alpha, \beta)^T\leftrightarrow(\beta,\alpha)^T$ simply corresponds to an exchange of the two modes and $\mathcal{T}: i \rightarrow -i$. The eigenvalues and (unnormalized) eigenvectors of $H$ are~\cite{Ramezani} $\lambda_{\pm}=\pm \sqrt{g^2 -  \Gamma^2}$, $\psi_{+}=( e^{i \frac{\theta}{2}}, e^{-i \frac{\theta}{2}})^T$ and $\psi_{-}=(i e^{-i \frac{\theta}{2}}, -ie^{i \frac{\theta}{2}})^T$, where $ \sin(\theta)=\Gamma/g$. We see that despite being non-Hermitian, for $\Gamma \leq g$ both eigenvalues $\lambda_\pm$ are real, indicating purely coherent oscillations between the two modes. In this regime of unbroken $\mathcal{PT}$ symmetry the eigenvectors are eigenstates of the symmetry operator, i.e., $\mathcal{PT}\psi_{\pm}=\psi_{\pm}$.  For $
\Gamma>g$, both eigenvalues become imaginary and correspond to a gain and a loss eigenmode. In this regime $\theta$ is complex and $\psi_{\pm}$ no longer possess the same symmetry as $H$. One thus speaks of a $\mathcal{PT}$-symmetry-breaking transition occurring at the (exceptional) point $\Gamma=g$~\cite{Bender2007}.  

This conventionally studied $\mathcal{PT}$-symmetry-breaking effect captures the change in the system's transient dynamics, as it is observed, e.g., in the propagation of coupled optical modes through an active medium with loss and gain~\cite{Guo2009, Rueter2010}. However, a simple eigenvalue analysis does not afford any conclusions concerning its steady state.
Firstly, due to an exponentially amplified mode in the symmetry-broken phase, nonlinear effects must be taken into account in order to determine the system's long-time behavior~\cite{Musslimani2008,ZezyulinPRL2012,LumerPRL2013,AlexeevaPRA,BarashenkovJPA2015}. Secondly, due to non-decaying oscillations in the $\mathcal{PT}$-symmetric phase, any source of noise would heat up the system indefinitely, and within a linearized model the steady state would again be ill-defined.


\section{Steady-state $\mathcal{PT}$-symmetry breaking}

\begin{figure}[t]
  \centering
    \includegraphics[width=0.9\textwidth]{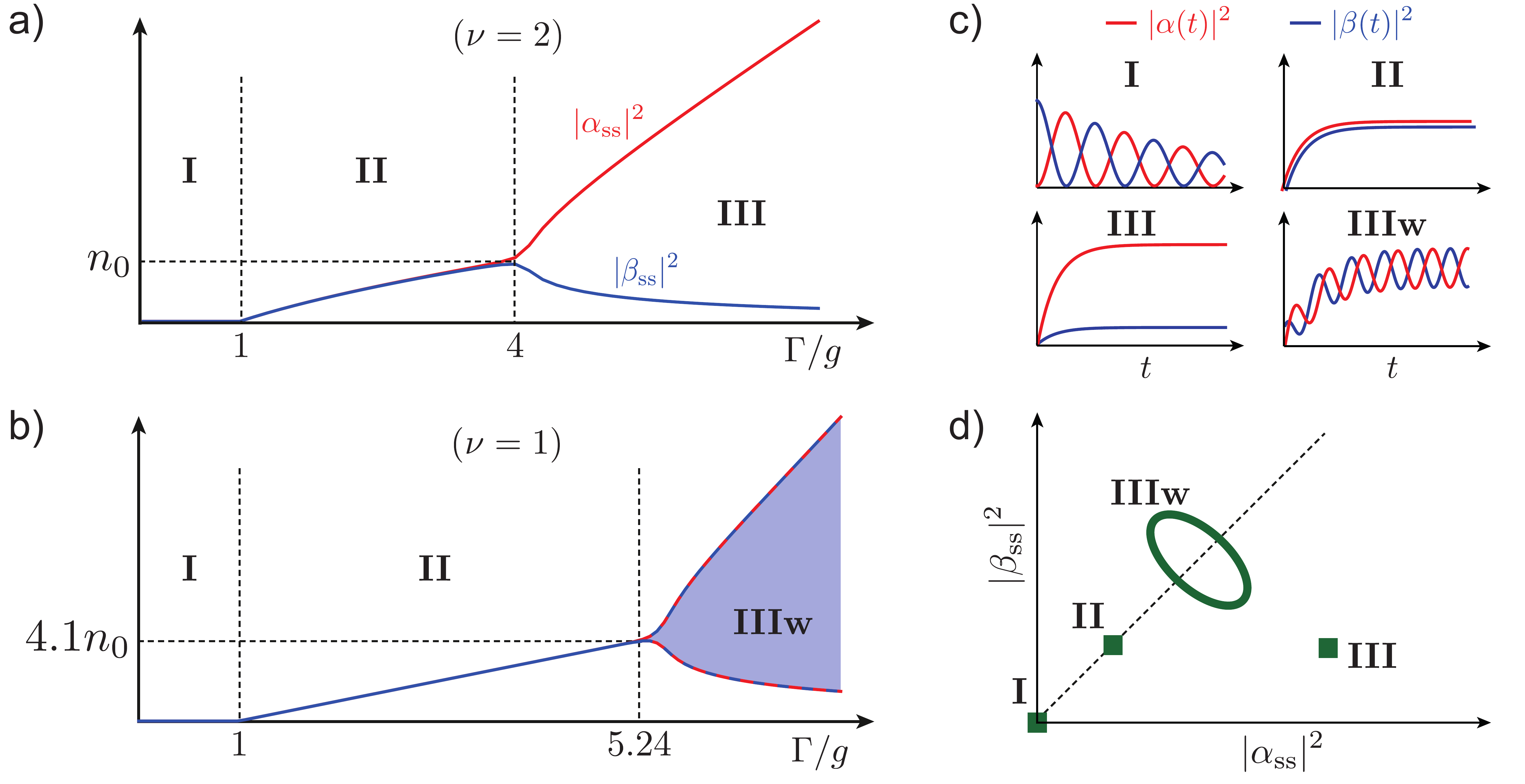}
      \caption{Steady state $\psi_{\rm ss}=(\alpha_{\rm ss},\beta_{\rm ss})^T$ of the $\mathcal{PT}$-symmetric phonon system in the absence of noise. Steady state occupation numbers of the two modes for a) $\nu = 2$ and b) $\nu = 1$ and $\gamma/g=10^{-3}$. In the limit-cycle phase ${\bf IIIw}$ both modes oscillate over the range indicated by the shaded area. c)~Illustration of the relaxation dynamics of $|\alpha|^2$ (red) and $|\beta|^2$ (blue). d)~The resulting steady state (green square) for each phase.}
      \label{phasesFig}
\end{figure}

Let us now return to the full model in equation~\eqref{stochasticEq} and evaluate the stationary state $\psi_{\rm ss} = (\alpha_{\rm ss}, \beta_{\rm ss})^T$ in the absence of noise. Figures~\ref{phasesFig}~a) and~b) show the mode occupation numbers $|\alpha_{\rm ss}|^2$ and $|\beta_{\rm ss}|^2$ as a function of $\Gamma/g$, and for the two relevant cases $\nu=2$ and $\nu=1$. Firstly, we observe in both plots the expected transition at $\Gamma/g |_{\bf I\rightarrow II}= 1$, below which (phase ${\bf I}$) the system dynamics is oscillatory with a small overall damping to $|\alpha_{\rm ss}|^2=|\beta_{\rm ss}|^2=0$ due to a finite $\gamma>0$.  Above this transition point (phase {\bf II}) the linearized system dynamics becomes unstable and both resonators reach a finite steady-state occupation number $|\alpha_{\rm ss}|^2/n_0 = |\beta_{\rm ss}|^2/n_0 =  (\Gamma/ g)^{1/\nu} -1 + \mathcal{O}(\gamma)$, determined  by the saturation of $\Gamma_\pm(\alpha)$. However, this steady state is still an eigenstate of the symmetry operator, $\mathcal{PT}\psi_{\rm ss}\propto \psi_{\rm ss}$, and contrary to our na{\" i}ve expectation the system remains $\mathcal{PT}$-symmetric beyond the conventional transition point.

The existence of a $\mathcal{PT}$-symmetric steady-state with non-vanishing amplitude is permitted in models like the present one, where $\mathcal{PT}$-symmetry is retained even in the nonlinear regime.   This means that a steady state $\psi_{\rm ss}$ with $|\alpha_{\rm ss}|=|\beta_{\rm ss}|\neq 0$ results in an \emph{equal} suppression of both the gain and the loss rate, $\Gamma_+(\alpha_{\rm ss})=-\Gamma_-(\beta_{\rm ss})$ [see equation~\eqref{nonlinearity}]. This implies that there exists a symmetric state which satisfies $\dot \psi_{\rm ss}=0$ for all values of $\Gamma/g$. However, as one can see in figure~\ref{phasesFig}, at larger values of  $\Gamma$ the system eventually switches to a different, symmetry-broken state. As we discuss now, the details of this $\mathcal{PT}$-symmetry-breaking mechanism and the resulting state depend on the actual form of the saturable gain, which is here determined by the parameter $\nu$.

Considering first the case $\nu=2$, we find a second transition at $\Gamma/g=4$, beyond which  (phase ${\bf III}$)
\begin{equation}\label{eq:SSIII}
\left.\begin{array}{c}
 |\alpha_{\rm ss}|^2 \\
|\beta_{\rm ss}|^2  
\end{array} \right\} 
=  n_0 \times\left(\frac{\Gamma \pm \sqrt{\Gamma (\Gamma - 4g)}}{2 g}-1\right)
\end{equation} 
and the steady state breaks parity, i.e., $|\alpha_{\rm ss}|\neq|\beta_{\rm ss}|$. The $\mathcal{PT}$-symmetry-breaking point can be obtained from a linear stability analysis for phase {\bf II} and for the present model we find \begin{equation}\label{eq:Transition1}
\left.\frac{\Gamma}{g}\right|_{{\bf II}\rightarrow{\bf III}}= \left(\frac{\nu}{\nu - 1} \right)^{\nu}. 
\end{equation} 
This result would imply the absence of symmetry breaking for $\nu=1$. Instead, for this case we observe a Hopf bifurcation around $\Gamma/g\approx 5.2$, beyond which the system approaches a limit cycle and undergoes periodic oscillations with the characteristic frequency $\omega_{\rm osc} \approx 2 \sqrt{g^3(\Gamma - g)} / \Gamma$.  The limit cycle is formed symmetrically around a central point $\psi_{\rm c}$ with $|\alpha_{\rm c}|=|\beta_{\rm c}|$, meaning that $\mathcal{PT}$-symmetry is preserved on average, but not at each point in time. To distinguish this behavior from the previous case we say that in this phase (${\bf IIIw}$) the $\mathcal{PT}$ symmetry  is only `weakly' broken.  In contrast to full symmetry breaking, this transition is related to the finite asymmetry induced by $\gamma>0$, and would be hidden in the idealized models where $\gamma=0$. The value for the transition point is
 \begin{equation}\label{eq:Transition2}
\left.\frac{\Gamma}{g}\right|_{{\bf II}\rightarrow{\bf IIIw}} =  \left(\frac{\nu + 2\nu^2 + \sqrt{2\nu + 3 \nu^2}}{ 2\nu^2-1} \right)^{\nu},
\end{equation}
independent of the precise value of $\gamma\ll\Gamma$.  The general expressions in equations~\eqref{eq:Transition1} and~\eqref{eq:Transition2} show that for all intermediate cases, $1<\nu<2$, we obtain $1< (\Gamma/g)_{{\bf II}\rightarrow{\bf IIIw}} < (\Gamma/g)_{{\bf II}\rightarrow{\bf III}} <\infty$ and $\mathcal{PT}$-symmetry breaking occurs in three steps ${\bf I}\rightarrow {\bf II}\rightarrow {\bf IIIw}\rightarrow {\bf III}$. More details on the derivation of the above results and the stationary phases for general $\nu$ are given in~\ref{sec:DynamicalAnalysis}.

While similar nonlinear phenomena are in general expected for gain-loss systems~\cite{Lu2015,AlexeevaPRA,PengScience2014,BrandstetterNatComm}, our specific interest here is to understand the role of dynamical instabilities in the breaking of a steady-state symmetry. In particular, the above analysis shows that $\mathcal{PT}$-symmetry breaking occurs even in systems where the symmetry is fully retained in the nonlinear regime and a symmetric steady-state would be permitted in principle for all parameters.  

\section{$\mathcal{PT}$-symmetry breaking in noisy systems}

We now show how the above picture changes in the presence of noise. For clarity we first restrict ourselves to a system which is dominated by thermal diffusion, i.e. $D_\pm = D=2\gamma N_{\rm th}$. For $\Gamma=0$ thermal noise induces additional amplitude fluctuations of about $(\Delta \alpha)^2\approx D/(2\gamma)=N_{\rm th}$, and we expect that as long as $N_{\rm th}<  n_0$ the characteristic features shown in figure~\ref{phasesFig}, which scale with the saturation number $n_0$, will only be smeared out, but not change significantly. This is confirmed numerically (not shown) and means that figure~\ref{phasesFig} is a good representation of the steady states of the system in the weakly nonlinear or low-noise regime. Therefore, we will now address the opposite regime $N_{\rm th}\gg n_0$.   

\begin{figure}[t]
  \centering
    \includegraphics[width=0.8\textwidth]{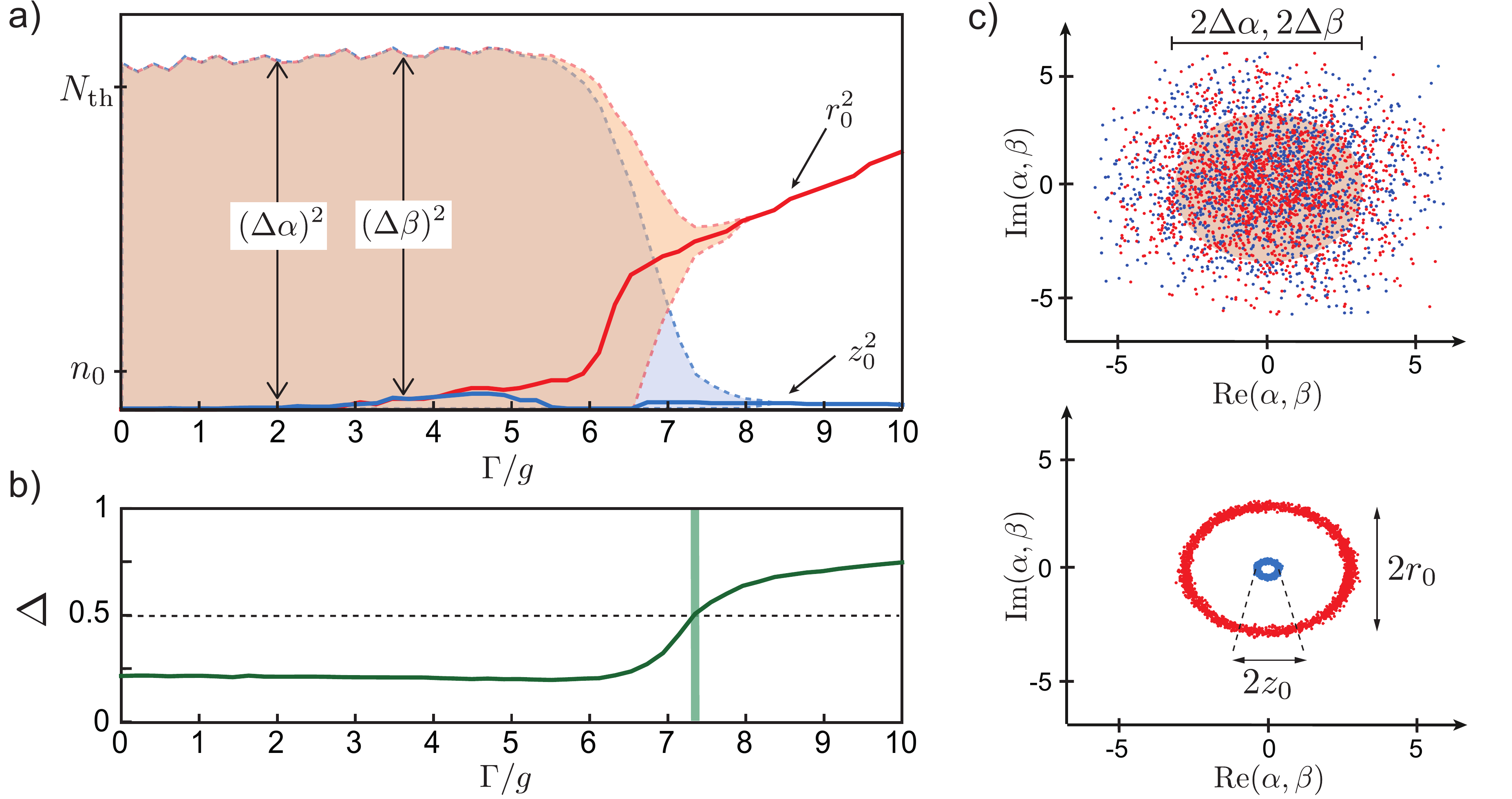}
      \caption{$\mathcal{PT}$-symmetry breaking in the limit of large thermal noise, $N_{\rm th}\gg n_0$. a) Steady-state distributions of $\alpha$ and $\beta$ for $N_{\rm th} / n_0=10$ and $\gamma/g=10^{-3}$. The values of $r_0$ and $z_0$ (solid lines) represent the radial distance of the distribution maxima from the origin and shaded areas indicate the range of fluctuations. b) Plot of the $\mathcal{PT}$-symmetry parameter $\Delta$ defined in equation \eqref{DeltaEq1}. c) Steady-state distribution of $\alpha$ (red dots) and $\beta$ (blue dots) in the thermal ($\Gamma/g=2$, left plot) and in the symmetry-broken ($\Gamma/g=10$, right plot) phases.}
      \label{noiseFig}
\end{figure}

Figure~\ref{noiseFig}~a) shows the results of a numerical simulation of the stochastic equation~\eqref{stochasticEq}, from which we obtain the steady-state distribution $P_{\rm ss}(\alpha,\beta)$ for $\nu=2$ and $N_{\rm th}/n_0=10$.  In the following we write $\alpha=r e^{i \theta_\alpha}$ and $\beta=z e^{i\theta_\beta}$ and make use of the fact that the system dynamics is invariant under a global phase rotation.  The exact marginal distributions $P_{\rm ss}(\alpha)$ and $P_{\rm ss}(\beta)$ are then fitted by approximate distributions of the form 
\begin{equation}
P(\alpha) \sim r   e^{-\frac{(r  - r_0)^2}{ \Delta \alpha^2}}, \qquad P(\beta ) \sim  z   e^{-\frac{(z  - z_0)^2}{ \Delta \beta^2}},
\end{equation}
which allows us to extract a radial shift $r_0$ and $z_0$ and the range of fluctuations $(\Delta \alpha)^2$ and $(\Delta \beta)^2$ for both modes (see~\ref{sec:Numerics}). From these values plotted in figure~\ref{noiseFig}~a) we see that the thermal noise now completely washes out the features associated with the $\mathcal{PT}$-symmetric phases {\bf I} and {\bf II}, and for a large range of $\Gamma$ the system reaches a steady state (phase ${\bf T}$), which is to a good approximation thermal, i.e., 
$r_{0}=z_0=0$ and $(\Delta \alpha)^2\simeq (\Delta \beta)^2\simeq N_{\rm th}$.
Only after a critical value of $\Gamma/g|_{\bf T\rightarrow III}\approx 7.5$ are the fluctuations suddenly strongly suppressed. In this regime the system relaxes into an asymmetric coherent state with $r_0> z_0$ approximately given by the amplitudes $|\alpha_{\rm ss} |$ and $|\beta_{\rm ss}|$ given in equation~\eqref{eq:SSIII} and  $(\Delta \alpha)^2,(\Delta \beta)^2\sim \gamma N_{\rm th}\Gamma /g^2, \gamma N_{\rm th}/\Gamma\ll1$. 

Before we proceed, let us connect this transition to the phenomenon of $\mathcal{PT}$-symmetry breaking---hitherto defined only for individual states. To do so we introduce the $\mathcal{PT}$-symmetry parameter
\begin{equation}\label{DeltaEq1}
\Delta   = \frac{ \langle (|\alpha| -|\beta|)^2\rangle_{\rm ss} }{ \langle |\alpha|^2\rangle_{\rm ss} +\langle|\beta|^2\rangle_{\rm ss}}\leq 1,
\end{equation}
which vanishes for a random set of states $\psi_i=(\alpha_i,\beta_i)^T$ if and only if each state satisfies  $\mathcal{PT}\psi_i = e^{i\theta_i}\psi_i$, with some real phase $\theta_i$. 
Figure~\ref{noiseFig}~b) shows that indeed $\Delta$ changes at the transition point from $\Delta\simeq \Delta_{\rm th}=0.215$ for a thermal state to $\Delta\rightarrow 1$ in the symmetry-broken phase. Note that also in the low-noise limit 
we obtain $\Delta=\Delta_{\rm th}>0$ in phase ${\bf I}$ and therefore only phase ${\bf II}$, where $\Delta \simeq 0$, has a strictly $\mathcal{PT}$-symmetric steady state.

One of the most striking features visible in figures~\ref{noiseFig}~a) and c) is that in sharp contrast to a conventional lasing transition,
 the emerging coherent-state amplitudes after the $\mathcal{PT}$-symmetry breaking point are even smaller than the original level of thermal noise. This surprising effect can be understood as follows. Although at each instant in time the amplitudes $\alpha(t)$ and $\beta(t)$, and therefore the gain and loss rates $\Gamma_+(\alpha)$ and $\Gamma_-(\beta)$, can be quite different, the average dissipation rate $\bar \Gamma =\langle \Gamma_--\Gamma_+\rangle_{\rm ss}$ when evaluated in the thermal phase vanishes, $\bar \Gamma \sim \mathcal{O}(\gamma)\sim 0$. What remains (on average) is the weak coupling to the high-temperature environment. In contrast, in the symmetry-broken phase we have $\langle |\alpha|\rangle_{\rm ss} \gg  \langle |\beta|\rangle_{\rm ss}$. Therefore, there is a strong imbalance between loss and gain on average, i.e., $\bar \Gamma \sim \mathcal{O}(\Gamma)>0$, and the resulting net cooling effect suppresses fluctuations.  Thus, this transition in the average dissipation rate of a stationary system can be seen as the counterpart of the transition from real to imaginary eigenvalues in the conventional definition of $\mathcal{PT}$-symmetry breaking. What we are still missing, however, is a simple criterion, which tells us why the system favors one or the other steady state. 

 \begin{figure}
  \centering
    \includegraphics[width=0.6\textwidth]{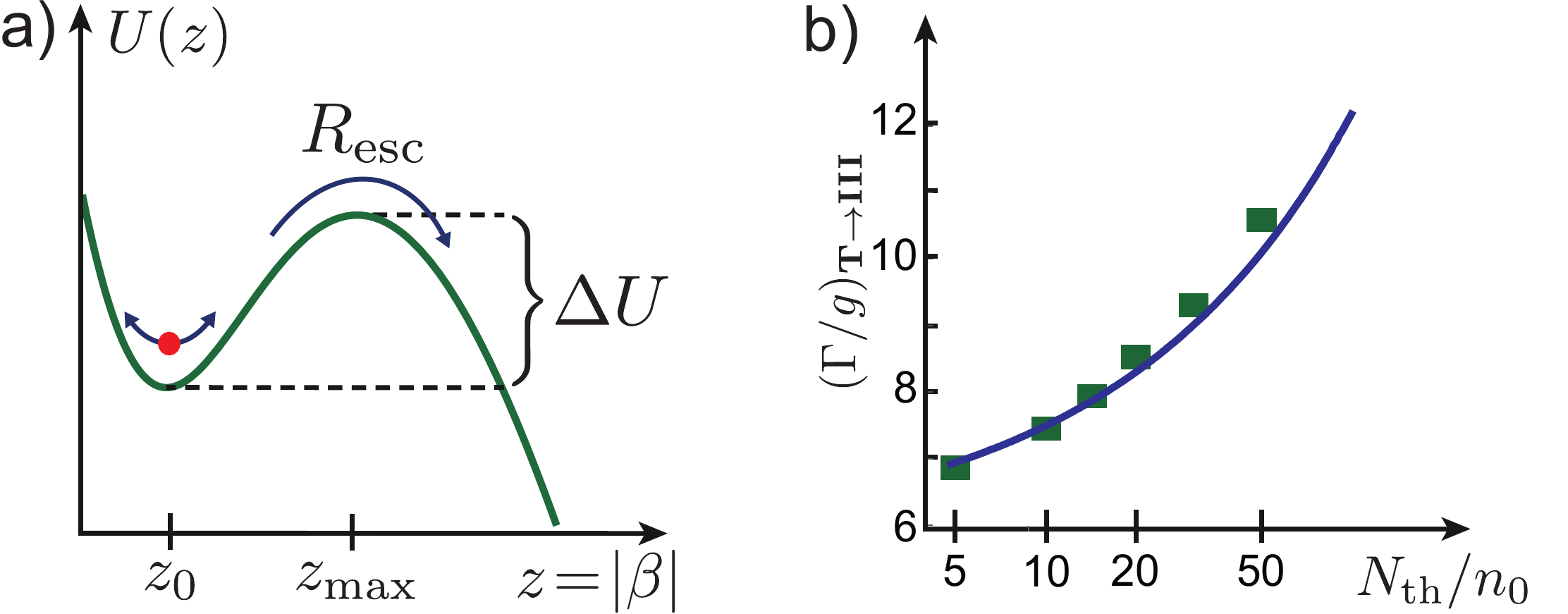}
      \caption{a)~Effective potential for the amplitude $z=|\beta|$ of the loss mode for $\Gamma/g=12$. b) Dependence of the symmetry breaking point  $\Gamma/g|_{\bf T\rightarrow III}$ on $N_{\rm th}$. The solid line represents the prediction from our analytic model and the squares show the transition points obtained from the condition $\Delta=0.5$ in numerical simulations.}
      \label{krammersFig}
\end{figure}

To clarify the mechanism behind the symmetry-breaking transition we focus on the symmetry-broken regime $\Gamma / g \gg 1$, where we can assume that the amplitude of the gain mode $|\alpha(t)| \approx |\alpha_{\rm ss}|$ and the relative phase $\phi \simeq \pi / 2$ are approximately constant. We then obtain the equation of motion for the amplitude of the loss mode, $z = |\beta|$~\cite{thermalSymmetry},  
\begin{equation}
\partial_t z= -\partial_z U(z) +\sqrt{\gamma N_{\rm th}} \eta_z(t),
\end{equation}
where $\langle \eta_z(t)\eta_z(t')\rangle=\delta(t-t')$ and (for $\gamma\rightarrow 0$)
\begin{equation}\label{potenEq}
U(z) = - \frac{n_0 \Gamma }{2(1+z^2/n_0)} - g |\alpha_{\rm ss}| z \sin(\phi)  -\frac{\gamma N_{\rm th}}{2} \log(z).
\end{equation}
The function $U(z)$ is an effective potential for $z$, which is sketched in Fig.~\ref{krammersFig}~a). This potential has a local minimum at $z_0=|\beta_{\rm ss}|$ (corresponding to the steady state given in equation~\eqref{eq:SSIII} for $N_{\rm th}\rightarrow 0$), which is separated by a finite barrier $\Delta U$ from the unstable region $z>z_{\rm max}$. In the presence of noise, a system initially located at $z\approx z_{0}$ can escape  over this barrier via thermally activated processes with a characteristic rate  $R_{\rm esc}\simeq R_0 e^{-\frac{2\Delta U}{\gamma N_{\rm th}}}$, where $R_0=\sqrt{-U''(z_{\rm min})U''(z_{\rm max})/(4\pi^2)}$~\cite{Gardiner}. This rate increases as $\Gamma$ is reduced and once $R_{\rm esc}$ exceeds the bare damping $\gamma$, any configuration with fixed $\alpha$ and $\beta$ is rapidly destabilized and the transition to a quasi-thermal state with strongly fluctuating amplitudes occurs. In figure~\ref{krammersFig}~b) we compare the transition point $ \Gamma/g|_{\bf T\rightarrow III}$ obtained from the condition $\gamma=R_{\rm esc}$ with the numerically evaluated values for various $N_{\rm th}/n_0\gg1$. The plot shows that $\mathcal{PT}$-symmetry breaking in the large-noise regime is very well described by this thermal activation model.

\section{Quantum noise limit}
As it has been pointed out above, the implementation of any gain mechanism is necessarily accompanied by a minimum amount of quantum noise, $D_{\rm q}$, which ensures, e.g., the preservation of Heisenberg's uncertainty relations in a quantum mechanical  amplification process. For $\mathcal{PT}$-symmetric systems this implies that noise is not only an intrinsic property, but also that the symmetry in the gain and loss modes is broken on a fundamental level. 
\begin{figure}[t]
  \centering
    \includegraphics[width=0.55\textwidth]{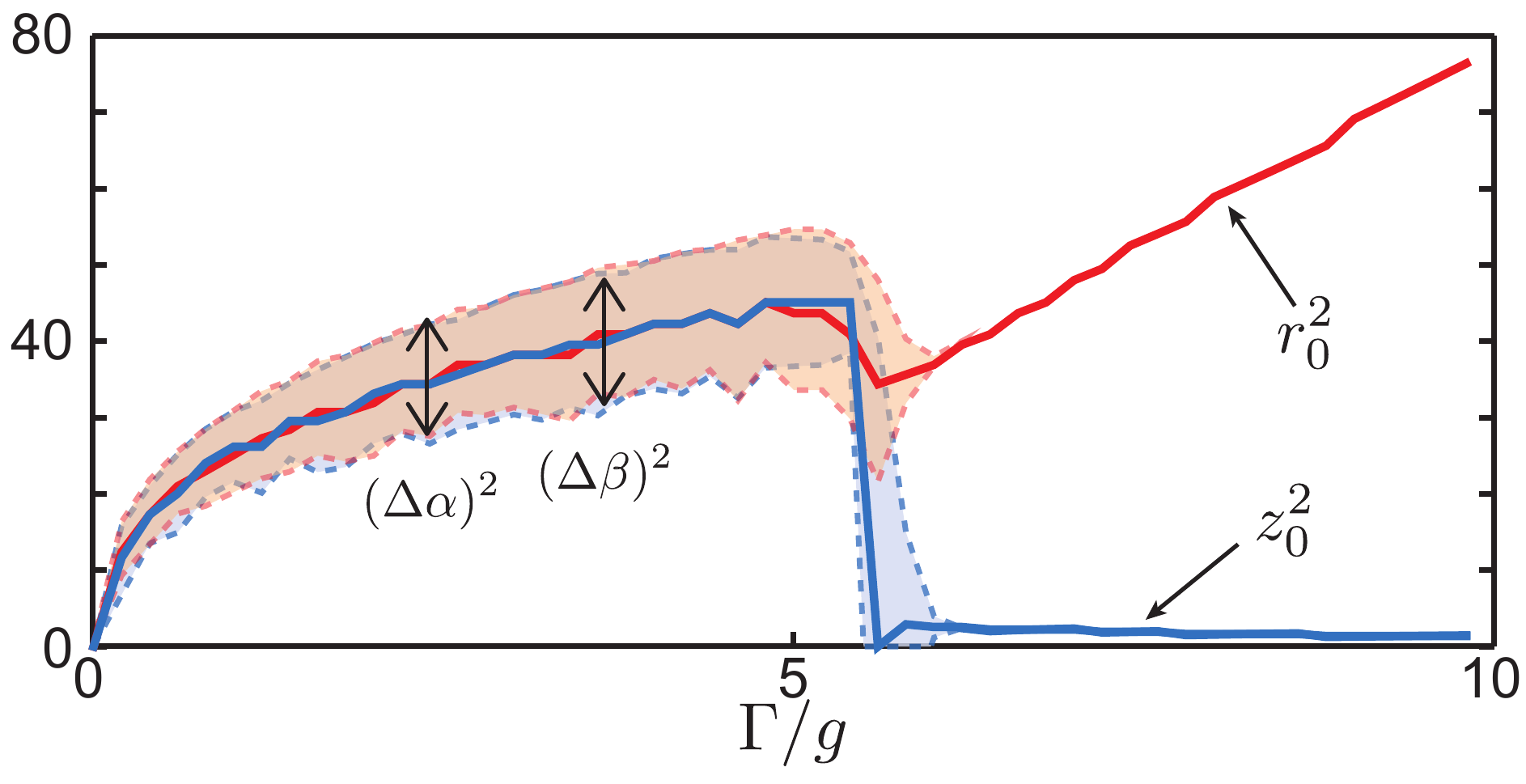}
      \caption{$\mathcal{PT}$-symmetry breaking in the quantum-noise limit. The radial position of the steady-state distribution maxima $r_0$ and $z_0$ (solid lines) and the range of fluctuations $\Delta \alpha$ and $\Delta \beta$ are plotted for the case $N_{\rm th}=0$ and $D_{\rm q}(\alpha)$ as defined in equation~\eqref{Dq}. The other parameters for this plot are $n_0=10$ and  $\gamma/g= 10^{-3}$. }
      \label{QuantumNoiseFig}
\end{figure}

To illustrate the effect of pure quantum noise on the system's steady state we set $N_{\rm th}=0$ and consider the quantum diffusion term
\begin{equation}\begin{split}\label{Dq}
  D_{\rm q}(\alpha) = \frac{2 \Gamma}{(1 + |\alpha|^2 / n_0)^3},
\end{split}\end{equation}
only. The assumed form of $D_{\rm q}(\alpha)$ applies for the three-level gain mechanism shown in figure~\ref{introFig}, i.e., for the case $\nu=2$ (see~\ref{sec:LaserTheory} for additional details). Similar to the thermal noise limit discussed above we represent the  resulting steady-state distributions for the loss and gain modes by the distribution maxima $r_0$, $z_0$ and the range of fluctuations $\Delta \alpha$ and $\Delta \beta$ respectively. The results are shown in figure~\ref{QuantumNoiseFig} for a saturation parameter  $n_0=10$. We see that while the steady-state distributions for low $\Gamma$ do no longer represent thermal states, there still exists a sharp transition between a high-noise $\mathcal{PT}$-symmetric phase and a $\mathcal{PT}$-symmetry-broken phase with strongly reduced fluctuations. Thus, we conclude that the $\mathcal{PT}$-symmetry-breaking mechanism identified above is not very sensitive to the details of the noise process and will exist in both thermal and quantum noise limited systems. Note that the semi-classical descriptions used in the current analysis is only valid for $n_0\gg1$ while for $n_0\sim 1$ where quantum effects are most important a full quantum mechanical treatment must be employed. Such an analysis is, however, beyond the scope of the current work and will be presented elsewhere.

\section{Arrays}
\begin{figure}
  \centering
    \includegraphics[width=0.6\textwidth]{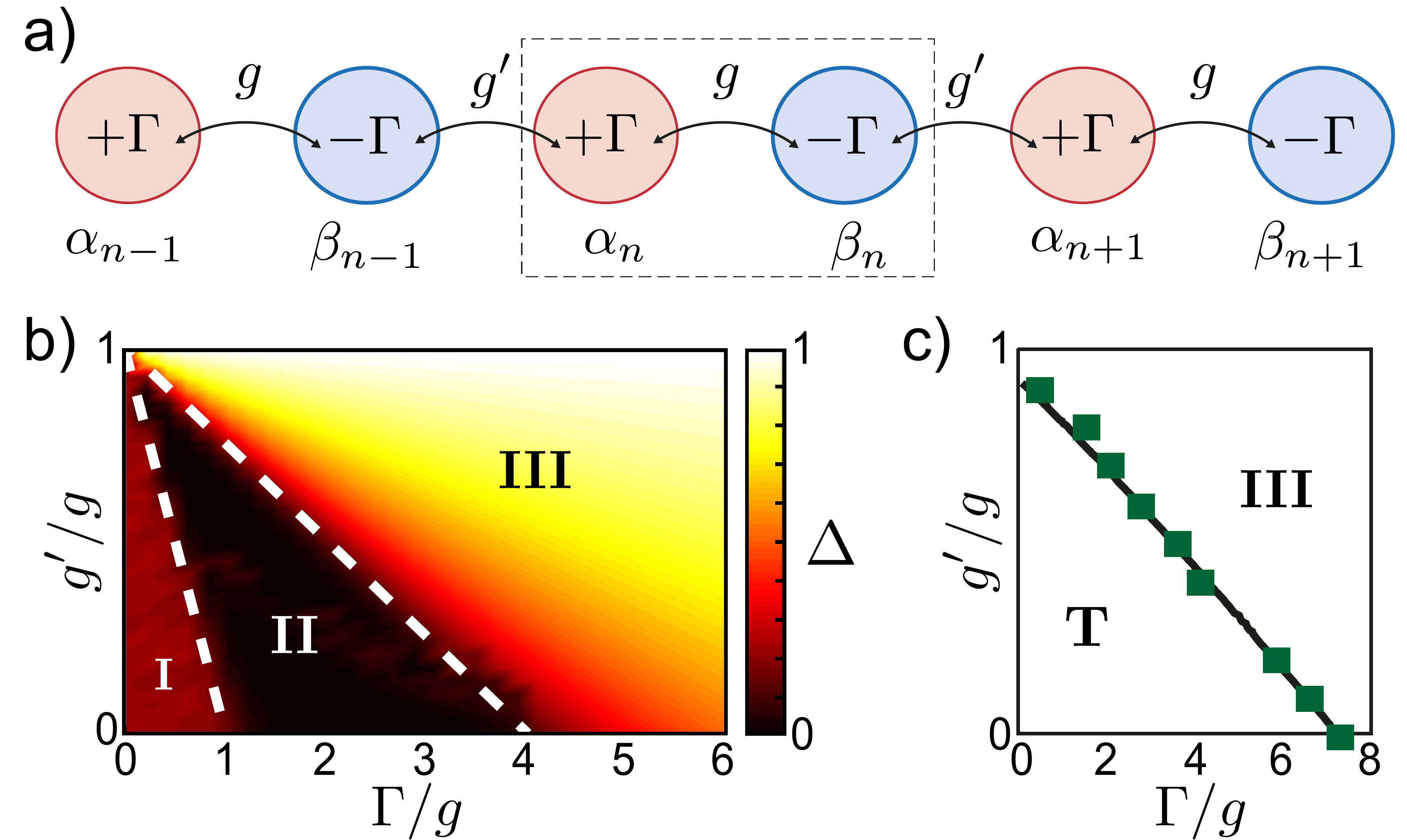}
      \caption{ a)~Schematic representation of an array of coupled resonators with a $\mathcal{PT}$-symmetric unit cell. b, c)~Steady state of a $\mathcal{PT}$-symmetric array of $N=12$ resonators in b)~the low noise regime, $N_{\rm th}/n_0=0.01$ and c)~the large noise limit $N_{\rm th}/n_0=10$. In~b) the different phases are characterized by the symmetry parameter $\Delta$ of a single unit cell and the white-dashed lines indicate the phase boundaries obtained from a plane wave ansatz. In~c) the solid line shows the analytic prediction for the phase boundary between the thermal and symmetry-broken phase and the green squares are the transition points obtained numerically. For both plots $\gamma/g=10^{-3}$ and $\nu=2$, such that the line  $g'=0$  corresponds to the setting considered in figures~\ref{phasesFig} a) and \ref{noiseFig}.}
      \label{arrayFig}
\end{figure}

Finally, to show that $\mathcal{PT}$-symmetry breaking in the steady state exists also for extended systems, we generalize the analysis above to coupled resonator arrays with alternating gain and loss~\cite{Lazarides,Vazquez-Candanedo}, as depicted in figure~\ref{arrayFig}~a). In this case the amplitudes $\alpha_n$ and $\beta_n$ for each unit cell obey
\begin{equation}\begin{split} \label{stochasticEqN}
\left( \begin{array}{c}
  \dot\alpha_n \\
\dot \beta_n  \end{array} \right) 
= & 
\left( \begin{array}{cc}
 \Gamma_+(\alpha_n)  & -i g  \\
-ig &  \Gamma_-(\beta_n)  \end{array} \right)
\left( \begin{array}{c}
  \alpha_n \\
\beta_n  \end{array} \right)  
 -i  g^{\prime} \left( \begin{array}{c}
 \beta_{n-1} \\
\alpha_{n+1} \end{array} \right)
+\left( \begin{array}{c}
 F_{n,+}(t) \\
F_{n,-}(t) \end{array} \right), 
\end{split}\end{equation}
where $g'$ is the coupling between the unit cells and $F_{n,\pm}(t)$ are independent thermal noise processes. Figure~\ref{arrayFig} summarizes the numerical results for the steady state of an array of $N=12$ resonators with periodic boundary conditions and $\nu = 2$. The observed features can be understood from the plane wave ansatz~\cite{Lazarides,CrossHohenbergRevModPhys} $\alpha_n = A_k e^{ik n}$, $\beta_n = B_k e^{ik n}$, where $k=  4 \pi j / N$. This ansatz maps equation~\eqref{stochasticEqN} onto a two-mode problem for $A_k$ and $B_k$, which is equivalent to equation~\eqref{stochasticEq}, but with the replacement  $g\mapsto g_k = |g + g' e^{ik}|$. This implies that the largest ratio $\Gamma/|g_k|$ is always achieved for the $k=\pi$ mode~\cite{Vazquez-Candanedo}, which therefore determines the symmetry-breaking properties of the array. For the special case $g=g'$, i.e., $g_k=0$, the gain and the loss modes completely decouple, and $\mathcal{PT}$-symmetry breaking already occurs for $\Gamma\rightarrow 0$~\cite{Feng2014}. For all intermediate parameters the phase boundaries in figure~\ref{arrayFig} are obtained from the analytic results for the two-mode problem, but with $g$ replaced by $g-g'$. We see that the single-mode ansatz captures well the relevant physics both in the low- and high-noise regime. Note that, however, in the `thermal' phase the behavior of the array can actually be much more complicated, since the system may undergo noise-induced jumps between multiple metastable configurations.

\section{Conclusions}

We have analyzed the breaking of $\mathcal{PT}$ symmetry in the steady state of coupled mechanical systems with loss and gain. We have shown how saturation effects and the influence of thermal or quantum noise lead to surprising new effects, which are not expected from the conventional eigenvalue analysis of idealized $\mathcal{PT}$-symmetric systems. These findings constitute a significant and important first step towards a general understanding of the behavior of realistic systems with $\mathcal{PT}$ symmetry, especially for systems operated close to the quantum regime or systems with strong nonlinearities.

\acknowledgements

We thank T.~E.~Lee, G.~J.~Milburn, and H.~E.~T{\" u}reci for stimulating discussions. This work was supported by the European FP7/ITC Project SIQS (600645), by the People Program (Marie Curie Actions) of the European Union under REA grant agreement number PIOF-GA-2011-303228 (project NOLACOME), project OPSOQI (316607) of the WWTF, the Austrian Science Fund (FWF) through SFB FOQUS F40, SFB NextLite F49-10, project GePartWave No.\@ I 1142-N27, DK CoQuS W 1210 and the START grant Y 591-N16. K.G.M. is also supported by the European Union Seventh Framework Programme (FP7-REGPOT-2012-2013-1) under grant agreement 316165.

\appendix

\section{Semi-classical laser theory}
\label{sec:LaserTheory}
In this section we outline the main steps in the derivation of the stochastic differential equation~\eqref{stochasticEq}. To do so we first consider an interaction between a single mechanical mode and a second auxiliary system, for example an optically pumped defect center~\cite{Kepesidis} or quantum dot~\cite{WilsonRaePRL2004}, which is used to engineer mechanical gain or loss. We assume a general coupling of the form
\begin{equation}
H_{\rm int} = \lambda \left(A a^{\dagger} + A^\dag a \right),
\end{equation}
where $a$ is the bosonic operator of the mechanical mode and $A$ is an operator of the auxiliary system.  In the absence of the coupling the auxiliary system relaxes with a rate $\Gamma_{\rm a}\gg \lambda $ into a steady state described by the density operator $\rho_{\rm a}^0$. This separation of timescales allows us to adiabatically eliminate the dynamics of the auxiliary system and derive an effective equation of motion for the mechanical mode. Following the standard treatment of semi-classical laser theory (see, e.g., section~9.3.2 of reference~\cite{GardinerZoller}), the result of such a calculation is a Fokker-Planck equation of the form
\begin{equation}\label{fpEq}
\dot{P}(\alpha,t) = \left[ - \left(  \frac{\partial}{\partial \alpha}\alpha + \frac{\partial}{\partial \alpha^*}\alpha^*  \right)  \Gamma(\alpha)  + \frac{\partial^2}{\partial \alpha^* \partial \alpha} D(\alpha) \right] P(\alpha ,t).
\end{equation}
Here $P(\alpha)$ defined via $\rho_{\rm m} = \int d^2\alpha \, P(\alpha) |\alpha \rangle \langle \alpha| $ is the Glauber-Sudarshan phase-space representation or $P$-distribution~\cite{GardinerZoller,WallsMilburn} of the mechanical density operator $\rho_{\rm m}$. In this equation the total gain function 
\begin{equation}\begin{split}
	&\Gamma(\alpha) = \Gamma_{\text{q}}(\alpha)- \gamma, \qquad \Gamma_{\text{q}}(\alpha) = - \frac{i \lambda \langle A \rangle_{\rm ss}}{\alpha},
\end{split}\end{equation}
and the total diffusion function
\begin{equation}\label{diffusionEq}\begin{split}
	&D(\alpha) = 2 \gamma N_{\text{th}} + D_{\text{q}}(\alpha) ,\qquad D_{\text{q}}(\alpha) = 2 \lambda^2 \Re \int_0^{\infty} d\tau \, \left(\langle  A^\dag (\tau) A \rangle_{\rm ss} -  |\langle A \rangle_{\rm ss} |^2\right),
\end{split}\end{equation}
are defined in terms of the unperturbed expectation value and correlation function of the operator $A$, where $\langle \dots \rangle_{\rm ss}={\rm Tr}_{\rm a}\{ \dots \rho_{\rm a}^0\}$. The damping rate $\gamma$ and diffusion rate $2 \gamma N_{\text{th}}$ encapsulate the effect of thermal noise from the environment, whereas $D_{\text{q}}(\alpha)$ is the intrinsic quantum noise associated with the gain process.

Equation~\eqref{fpEq} is a standard Fokker-Planck equation and as such it is equivalent to an It\^{o} stochastic differential equation with drift $ \Gamma(\alpha)$ and diffusion $D(\alpha)$~\cite[section~4.3.5]{Gardiner}.  That is, the dynamics of the mechanical system may be modeled using the It\^{o} stochastic differential equation thus
\begin{equation}\label{eq:stochasticEq}
	d\alpha = \Gamma(\alpha) \alpha dt + \sqrt{D(\alpha)} dW_{\mathbb{C}},
\end{equation}
where $dW_{\mathbb{C}} = 2^{-1/2} (dW_1 + i dW_2)$ is a complex Wiener increment; $dW_1$ and $dW_2$ are independent standard Wiener processes~\cite[section~3.8.5]{Gardiner}. By extending this analysis to two modes with mutual coupling $g$ we then obtain the two coupled equations
\begin{align}
	d\alpha  &=  \Gamma_+(\alpha) \alpha dt  - i g \beta dt +\sqrt{D_+(\alpha)} dW_{\mathbb{C}, +},  \label{eq:coupledSDEs1}\\
	d\beta  &=  \Gamma_-(\beta)  \beta dt  -i g \alpha dt +\sqrt{D_-(\beta)} dW_{\mathbb{C}, -} .\label{eq:coupledSDEs2}
\end{align}
By formally introducing the white-noise forces $F_\pm(t)=dW_{\mathbb{C}, \pm}/dt$, equations~\eqref{eq:coupledSDEs1} and \eqref{eq:coupledSDEs2} reproduce equation~\eqref{stochasticEq}.

\subsection{Drift and diffusion for a phonon laser induced by a three-level defect}
\label{sec:three-levelSystem}

Reference~\cite{Kepesidis} considers a nitrogen-vacancy (NV) defect centre, which is coupled via strain to the bending motion of a diamond nanobeam. The NV center is modeled as a three-level system with a ground state $|g\rangle$ and two near-degenerate electronically excited states $|x\rangle$ and $|y\rangle$. If the excited-state splitting $\Delta E_{xy}$ matches the vibration frequency of the phonon mode $\omega_{\rm m}$ then resonant transitions are induced between the two states. By optically exciting the upper level $|x\rangle$ this process excites the phonon mode (i.e., effects gain) whereas optically exciting the lower level $|y\rangle$ leads to a cooling process (see Fig.~1~c) in the main text). In the gain configuration the system can be described by the Hamiltonian $(\hbar=1)$
\begin{equation}
H = \omega_m a^\dag a - \Delta E_{xy} |y\rangle\langle y|  + \frac{\Omega}{2} \left(|g\rangle\langle x| + |x\rangle\langle g|\right) + \lambda \left( |y\rangle\langle x| a^\dag + a |x\rangle\langle y| \right),
\end{equation}
where $\lambda$ is the strain coupling constant. For the cooling configuration we obtain the same type of coupling, but with the role of $|x\rangle$ and $|y\rangle$ reversed. By assuming a radiative decay rate $\Gamma_{\rm a}$ for both of the two excited states, $\omega_{\rm m}=\Delta E_{xy}$ (gain configuration) and weak driving $\Omega\ll \Gamma_{\rm a}$ we obtain~\cite{Kepesidis}
\begin{equation}\begin{split}
	&\Gamma_{\rm q} (\alpha) =  \frac{ \Gamma}{(1  +  |\alpha|^2/n_0)^2}, \qquad   D_{\rm q}(\alpha) = \frac{2 \Gamma}{(1 + |\alpha|^2 / n_0)^3},
\end{split}\end{equation}
where $\Gamma = 2 \lambda^2 \Omega^2 /\Gamma_{\rm a}^3$ and $n_0 = \Gamma_{\rm a}^2 / 4\lambda^2$. 

\section{Stability analysis in the absence of noise}
\label{sec:DynamicalAnalysis}

In this section we outline the stability analysis used to obtain the steady-state phases of two coupled resonators in the absence of noise, i.e., with $D_\pm=0$. By moving to polar coordinates $\alpha =  r e^{i \theta_\alpha}$ and $\beta = z e^{i \theta_\beta}$, equation~\eqref{stochasticEq} can be rewritten as
\begin{align}
	\dot{r} &= - [\gamma - \Gamma(r)] r - g \sin(\phi) z ,\label{eq:simpleSystem1}\\
		\dot{z} &= - [\gamma + \Gamma(z)] z + g \sin(\phi) r, \label{eq:simpleSystem2}
\\
	\dot{\phi} &= g \left(\frac{r}{z} - \frac{z}{r}\right) \cos(\phi),
\end{align}
where $\phi = \theta_\alpha - \theta_\beta$.  Note that the system is invariant under a combined rotation of $\alpha$ and $\beta$ and therefore the evolution of the total phase $\theta_\alpha+\theta_\beta$ can be neglected. For the last equation we see that there are two fixed points for the phase, $\phi_{\text{ss}} = \pm \pi / 2$.  Due to finite $\gamma$, the stationary occupation number of the gain mode $|\alpha|^2$ is always slightly larger than that of the loss mode $|\beta|^2$, wherefore $\phi_{\text{ss}} = \pi / 2$ is the stable.  We therefore set $\phi = \pi / 2$ and henceforth study the two-dimensional dynamical system with variables $r$ and $z$.

\begin{figure*}
  \centering
    \includegraphics[width=0.9\textwidth]{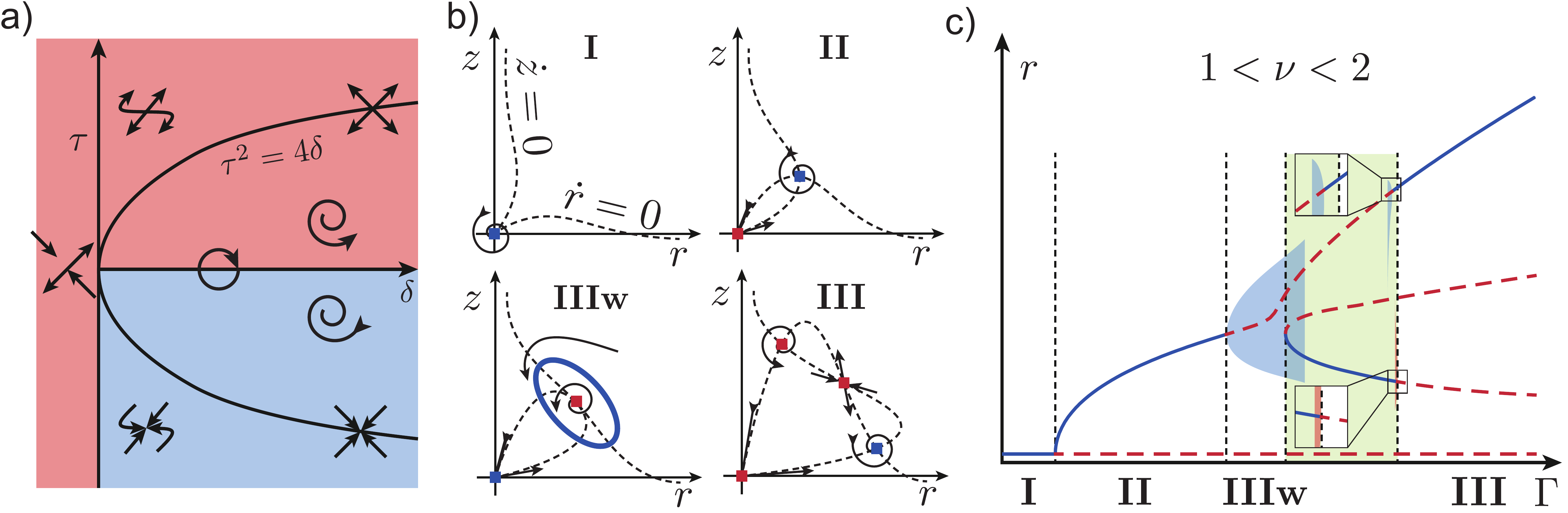}
      \caption{a)~Trace-determinant plane of the Jacobian for a two-dimensional dynamical system with cartoons of the local phase portraits (reproduced from~\cite[figure~5.2.8]{Strogatz}).  The blue- and red-shaded areas indicate stable and unstable regions respectively.  b)~Cartoons of the phase portraits for the four different phases.  The dashed lines denote nullclines (the curves in the $r$-$z$-plane specified by setting either $\dot{r} = 0$ or $\dot{z} = 0$; the intersections of nullclines correspond to stationary states), and the solid lines with arrows denote example integral curves (actual trajectories in $r$ and $z$).  c)~Cartoon of the general bifurcation diagram for $r$ for $1<\nu <2$.  Solid blue lines denote stable stationary states, dashed red lines unstable stationary states, and the shaded blue and red regions denote stable and unstable limit cycles respectively.  This plot has been made using data produced by the software MatCont~\cite{matcont} and we have chosen $\nu = 1.5$ and $\gamma / g = 0.001$.  We note the presence of intermediate phases in the transition from phase \textbf{IIIw} to phase \textbf{III}, the green-shaded region.  In this region the stable stationary state is such that $r < z$, which vitiates our assumption $|\alpha| > |\beta|$ that we used to eliminate the phase; for $|\alpha| < |\beta|$ the phase $\phi = \pi / 2$ is unstable.  We therefore expect some kind of oscillation in this intermediate regime, which is also observed in numerical simulations.}
      \label{fig:figure1}
\end{figure*}

In the following we apply standard dynamical analysis to understand the phases mentioned above and predict the transition points.  
To do so we first evaluate the possible stationary solutions $r_{\rm ss}$ and $z_{\rm ss}$ of equations~\eqref{eq:simpleSystem1} and~\eqref{eq:simpleSystem2}, which are given by the solutions of 
\begin{align}
	 g z_{\rm ss}=  [ \Gamma(r_{\rm ss})-\gamma] r_{\rm ss} ,\qquad g  r_{\rm ss}=  [ \Gamma(z_{\rm ss})+\gamma] z_{\rm ss}.
\end{align}
The stability of these fixed points is then analyzed using the trace-determinant plane of the Jacobian, i.e. the dynamical matrix of the system linearized about said stationary state.  The Jacobian for our system is
\begin{equation}
	\textbf{J} = \begin{pmatrix}
		\Gamma'(r_{\rm ss}) r_{\rm ss} + \Gamma(r_{\rm ss}) - \gamma & - g \\
		g & - \Gamma'(z_{\rm ss}) z_{\rm ss} - \Gamma(z_{\rm ss}) - \gamma
	\end{pmatrix},
\end{equation}
where the prime denotes the derivative.  The trace and determinant of $\textbf{J}$ are
\begin{gather}
	\tau = \operatorname{Tr} \textbf{J} = - 2 \gamma + \Gamma'(r_{\rm ss}) r_{\rm ss} - \Gamma'(z_{\rm ss}) z_{\rm ss} + \Gamma(r_{\rm ss}) - \Gamma(z_{\rm ss}) \label{eq:trace}, \\
	\delta = \operatorname{det} \textbf{J} = g^2 - [\Gamma'(r_{\rm ss}) r_{\rm ss} + \Gamma(r_{\rm ss}) - \gamma] [\Gamma'(z_{\rm ss}) z_{\rm ss} + \Gamma(z_{\rm ss}) + \gamma] \label{eq:det}
\end{gather}
respectively.  Since the eigenvalues of $\textbf{J}$ may be written entirely in terms of $\tau$ and $\delta$ thus
\begin{equation}
	\lambda_\pm = {\textstyle\frac{1}{2}} (\tau \pm \sqrt{\tau^2 - 4 \delta}) ,
\end{equation}
evaluating $\tau$ and $\delta$ at a particular stationary state fully characterizes its stability and local dynamical structure.  For example, if the real part of both $\lambda_+$ and $\lambda_-$ is negative then the stationary state considered is stable, and a non-vanishing imaginary part effects a curl in the phase portrait (the $r$-$z$-plane).  We summarize this characterization in figure~\ref{fig:figure1}~a).

\subsection{Phases}

\subsubsection*{Phase \textbf{I}:} The state $r_{\text{ss}} = z_{\text{ss}} = 0$ is an obvious stationary state of equations~\eqref{eq:simpleSystem1} and~\eqref{eq:simpleSystem2}.  Substituting this into equations~\eqref{eq:trace} and~\eqref{eq:det} yields (to first order in $\gamma$)
\begin{equation}\begin{split}
	&\tau = - 2 \gamma, \qquad \delta = g^2 - \Gamma^2 .
\end{split}\end{equation}
Consulting figure~\ref{fig:figure1}~a), we see that this corresponds to a stable spiral for $\Gamma < g$, but for $\Gamma > g$ becomes a saddle node.  We thus conclude that phase \textbf{I} is exhibited for
\begin{equation}
	0 < \frac{\Gamma}{g} < 1 .
\end{equation}

\subsubsection*{Phase \textbf{II}:} In phase \textbf{II} the two occupation numbers are roughly equal, however simply assuming $r_{ \text{ss}} = z_{ \text{ss}}$ yields an inconsistency due to finite $\gamma$.  Let us therefore substitute the ansatz $r_{\text{ss}}= r_{\text{ss}}^{(0)} + \gamma r_{\text{ss}}^{(1)} + \ldots$ and $z_{ \text{ss}}= r_{\text{ss}}^{(0)} + \gamma r_{\text{ss}}^{(1)} + \ldots$ and equate alike-orders in $\gamma$.  The result is
\begin{equation}\begin{split}\label{eq:stateII}
	&r_{\text{ss}}^{(0)} = \sqrt{n_0} \times \sqrt{(\Gamma / g)^{1/\nu} - 1}, \\
	& r_{ \text{ss}}^{(1)} = - z_{ \text{ss}}^{(1)} = r_{\text{ss}}^{(0)} [\Gamma'(r_{\text{ss}}^{(0)}) r_{\text{ss}}^{(0)} + \Gamma(r_{\text{ss}}^{(0)}) + g]^{-1} .
\end{split}\end{equation}
We neglect terms of order $\gamma^2$ and higher.  Note that, this stationary state only exists for $\Gamma / g \geq 1$.  Substituting equation~\eqref{eq:stateII} into equations~\eqref{eq:trace} and~\eqref{eq:det} yields
\begin{align}
	\tau &= - 2 \gamma \left(
		1 - \frac{\Gamma''(r_{\text{ss}}^{(0)}) r_{\text{ss}}^{(0)2} + 2 \Gamma'(r_{\text{ss}}^{(0)}) r_{\text{ss}}^{(0)}}
		{\Gamma'(r_{\text{ss}}^{(0)}) r_{\text{ss}}^{(0)} + \Gamma(r_{\text{ss}}^{(0)}) + g}
	\right) \\
	&= - 2 \gamma \left( 1 - \frac{2 \nu [(\Gamma / g)^{1 / \nu} - 1] [(2 \nu - 1) (\Gamma / g)^{1 / \nu} - 2 (\nu + 1)]}{2 \nu (\Gamma / g)^{1 / \nu} + 2 (1 - \nu) (\Gamma / g)^{2 / \nu}} \right)
	 \end{align}
and
\begin{equation}\begin{split}
	\delta &= g^2 - [\Gamma'(r_{\text{ss}}^{(0)}) r_{\text{ss}}^{(0)} + \Gamma(r_{\text{ss}}^{(0)})]^2  = g^2 - g^2 [2 \nu - 1 - 2 \nu (\Gamma / g)^{- 1 / \nu}]^2 .
\end{split}\end{equation}
We see that $\delta > 0$ only for $1 < \Gamma / g < [\nu / (\nu - 1)]^\nu$.  However, unlike the previous analysis of phase \textbf{I}, in this case $\tau$ can be positive whilst $\delta$ is positive.  Such is the case if $[\nu / (\nu - 1)]^\nu > [(\nu + 2 \nu^2 + \sqrt{2 \nu + 3 \nu^2}) / (2 \nu^2 - 1)]^{\nu}$.  Consulting figure~\ref{fig:figure1}~a), one sees that as $\Gamma / g$ is increased the stationary state~\eqref{eq:stateII} goes from a stable spiral to an unstable spiral and then to a saddle node.  We thus conclude that phase \textbf{II} is exhibited for
\begin{equation}
	1 < \frac{\Gamma}{g} < \operatorname{min} \left\{ \left( \frac{\nu + 2 \nu^2 + \sqrt{2 \nu + 3 \nu^2}}{2 \nu^2 - 1} \right)^{\nu} , \left( \frac{\nu}{\nu - 1} \right)^\nu \right\} .
\end{equation}

\subsubsection*{Phases ${\bf IIIw}$ and ${\bf III}$:} In both phase ${\bf IIIw}$ and ${\bf III}$ the occupation numbers are not even roughly equal. To analyze this regime it is instructive to consider the nullclines (the curves in the $r$-$z$-plane specified by setting either $\dot{r} = 0$ or $\dot{z} = 0$; the intersections of nullclines correspond to stationary states):
\begin{gather}
 z = f(r) - \frac{\gamma}{g} r ,
	\qquad r = f(z) + \frac{\gamma}{g} z, \qquad  f(r) = \frac{\Gamma}{g} \frac{r}{(1 + r^2/n_0)^\nu} .
\end{gather}
The function $f(r)$ is always positive, and only zero at $r = 0$ and $r \rightarrow \infty$.  The maximum is located at $r/\sqrt{n_0} = (2 \nu - 1)^{-1/2}$ where $f(r) = (\Gamma / g) (2 \nu)^{-1} (2 \nu - 1)^{\nu + 1/ 2}$.  The additional terms $-(\gamma / g) r$ and $(\gamma / g) z$ approximately rotate the $r$- and $z$-axes for $\gamma \ll g$.  The possible intersections of the two nullclines as a function of $\Gamma / g$ are shown in figure~\ref{fig:figure1} c): one intersection (\textbf{I} in the figure), two intersections (\textbf{II} and ${\bf IIIw}$ in the figure), or four intersections (\textbf{III} in the figure).  It is clear that phase ${\bf IIIw}$ corresponds to the case of two intersections, and from the previous section we know that if $[(\nu + 2 \nu^2 - \sqrt{2 \nu + 3 \nu^2}) / (2 \nu^2 - 1)]^{\nu} < \Gamma / g < [\nu / (\nu - 1)]^\nu$ then neither of these two intersections corresponds to a stable stationary state.  Therefore, by the Poincar{\'e}-Bendixon theorem~\cite[section~7.3]{Strogatz}, a limit cycle must exist.  We thus conclude that phase ${\bf IIIw}$ is exhibited for
\begin{equation}
	\left( \frac{\nu + 2 \nu^2 + \sqrt{2 \nu + 3 \nu^2}}{2 \nu^2 - 1} \right)^{\nu} < \frac{\Gamma}{g} < \left( \frac{\nu}{\nu - 1} \right)^\nu ,
\end{equation}
but the limit cycle persists slightly beyond this upper bound.  Since it is only possible that $[\nu / (\nu - 1)]^\nu > [(\nu + 2 \nu^2 + \sqrt{2 \nu + 3 \nu^2}) / (2 \nu^2 - 1)]^{\nu}$ if $1 \leq \nu < 2$, this phase cannot be observed for $\nu \geq 2$.  The limit cycle does not admit a simple analytic form, however by assuming that it is small and centered on stationary state~\eqref{eq:stateII}, which is the only unstable spiral for this regime, one may approximate its frequency $\omega_{\text{osc}}$ by the imaginary part of the eigenvalue of the Jacobian evaluated at equation~\eqref{eq:stateII}.  One finds $\omega_{\text{osc}} \approx 2 \sqrt{g^3 (\Gamma - g)} / \Gamma$; this result has been numerically verified for the case $\nu = 1$ (not shown).  On the other hand, it is also clear that phase \textbf{III} corresponds to the case of four intersections.  One may show that there are four intersections only if $\Gamma / g > [\nu / (\nu - 1)]^\nu$, and it appears that (at least for $\nu \geq 2$) of the two extra stationary states one must be an unstable spiral and the other a stable spiral.  We thus conclude that phase \textbf{III} is exhibited for
\begin{equation}
	\frac{\Gamma}{g} > \left( \frac{\nu}{\nu - 1} \right)^\nu .
\end{equation}
Note that this phase cannot be observed for $\nu = 1$.  For $\nu = 2$ one may easily check that, given $\Gamma / g > [\nu / (\nu - 1)]^\nu = 4$, the stable stationary amplitudes are 
\begin{gather}
	r_{ \text{ss}}/\sqrt{n_0} = \{[\Gamma + \sqrt{\Gamma (\Gamma - 4 g)}] / (2 g) - 1\}^{1/2} ,\\
	 z_{ \text{ss}}/\sqrt{n_0} = \{[\Gamma - \sqrt{\Gamma (\Gamma - 4 g)}] / (2 g) - 1\}^{1/2}.
\end{gather}

\section{Numerical simulations}
\label{sec:Numerics}

\begin{figure*}
  \centering
    \includegraphics[width=0.7\textwidth]{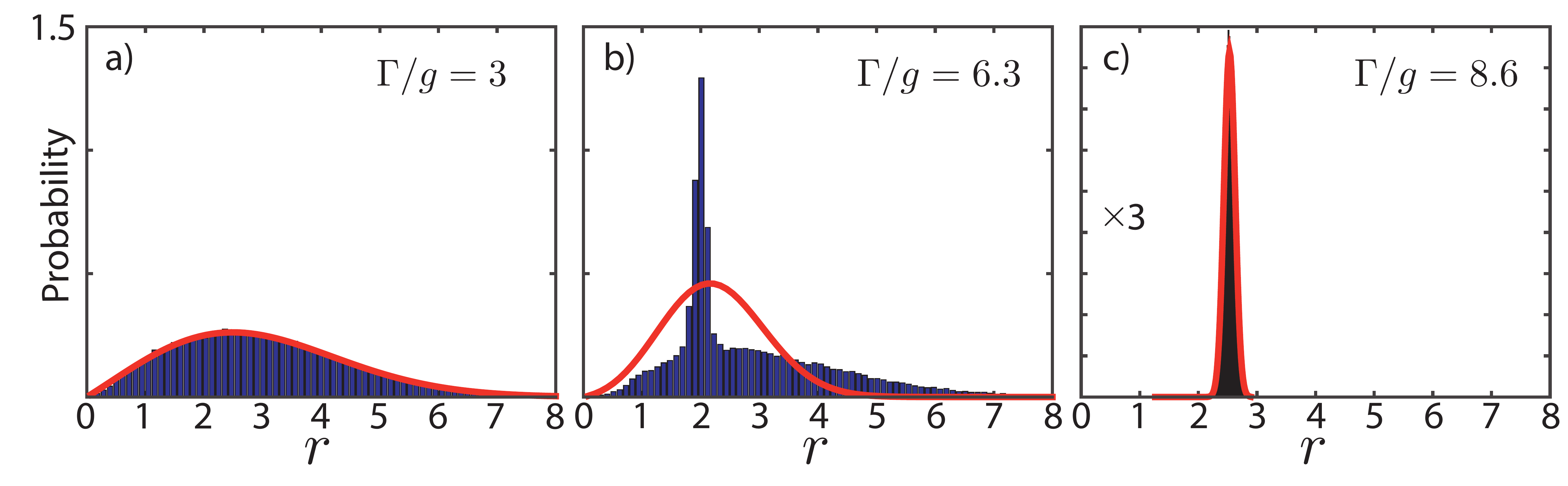}
      \caption{Histograms of the real amplitude $r = |\alpha|$ of the resonator with gain and the fitting distribution (red line)---described by equation~\eqref{distributionEq}---for three different values of the ratio $\Gamma/g$. a) Fitting distribution in the thermal regime, for $\Gamma/g = 3$. b) Fitting distribution in the vicinity of the thermally-activated transition, for $\Gamma/g = 6.3$. c) Fitting distribution in the lasing regime, for $\Gamma / g = 8.6$. Note that the $y$-axis of the last plot has been scaled by $1/3$.}
      \label{fig:figure2}
\end{figure*}

For the numerical results presented in Figs.~3 and~4 in the main text we have simulated the stochastic equations~\eqref{eq:coupledSDEs1} and \eqref{eq:coupledSDEs2} (and the corresponding extended set of equations for the array) using the Euler-Maruyama method (see section 15 of \cite{Gardiner}). Since we are interested in the steady state of the system, we collect data (complex numbers $\alpha(t)$ and $\beta(t)$) after a time of $~ 5 \times \tau_0$ where $\tau_0 = 1/\gamma$ is approximately the time scale in which the steady state is approached. Over a period of $ 45 \times \tau_0$ we select $4000$ random data points. We repeat the procedure $ 80 $ times with random initial conditions.

From the numerical data we obtain the marginal steady-state distributions $P_{\rm ss}(\alpha)$ and $P_{\rm ss}(\beta)$ for the two modes. Since the system is invariant under a combined rotation of $\alpha$ and $\beta$ in phase space, also the marginal distributions are radially symmetric. Figure~\ref{fig:figure2} shows examples of the resulting radial distribution $P_{\rm ss}(r=|\alpha|)$ for the gain mode, before, close to, and after the transition.  

To obtain a simple characterization of the system's steady state, we fit the radial distribution $P_{\rm ss}(r=|\alpha|)$ by a distribution of the form
\begin{equation}\label{distributionEq}
P_{r_0,\sigma}(r) = \mathcal{N} \times  r \times  e^{-\frac{(r - r_{0})^2}{ \sigma^2}},
\end{equation}
where $\mathcal{N}$ is a normalization constant. It corresponds to a thermal state for $r_0=0$ and $\sigma=\sqrt{N_{\rm th}}$ and approaches the distribution of a coherent state with random phase in the limit $\sigma\ll r_0$. The optimal values for $r_{0}$ and $\sigma$ are obtained by minimization of the squared difference
\begin{equation}
\chi^2(r_0,\sigma) = \sum_{i=1}^N \left(  H_i - P_{r_0,\sigma}(r_i)  \right)^2,
\end{equation}
where $H_i$ represents the hight of the $i$'th bar of the histogram. 

While this fitting procedure yields accurate values for the radial displacement of the distribution maximum, $r_0$, we find that the corresponding values for the width $\sigma$ do not very well capture the broad thermal background in the vicinity of the transition [see 
figure~\ref{fig:figure2} b)]. Therefore,  we use instead the quantity  
\begin{equation}
 (\Delta \alpha)^2 = \langle | \alpha |^2 \rangle - r_0^2,
 \end{equation}
where the average $\langle | \alpha |^2 \rangle$ is calculated directly from the data set, to represent the range of fluctuations.

\section*{References}

\end{document}